# Morphological tracking and tuning of silica NPs for stable levitation in vacuum optomechanical systems

Cuihong Li[1], Yuanyuan Ma[3], Jinchuan Wang[1], Shaochong Zhu[1], Xiaowen Gao[1], Xinbing Jiang[4], Huan Yang[5], Zhenhai Fu[1], Huizhu Hu[1,2,*]

[1] Research Center for Quantum Sensing, Intelligent Perception Research Institute, Zhejiang Lab, Hangzhou 310000, China
[2] State Key Laboratory of Modern Optical Instrumentation, College of Optical Science and Engineering, Zhejiang University, Hangzhou 310027, China
[3] The State Key Lab of Analytical Chemistry for Life Science, School of Chemistry and Chemical Engineering, Chemistry and Biomedicine Innovation Center (ChemBIC), Nanjing University, Nanjing 210093, China
[4] School of Chemistry, Engineering Research Center of Energy Storage Materials and Devices, Ministry of Education, Xi'an Jiaotong University, Xi'an 710049, China
[5] School of Physics, Xidian University, Xi'an 710071, China

(E-mail: huhuizhu2000@zju.edu.cn)

## Abstract

Optically levitated nanomechanical resonators in vacuum perform ultra-high sensitivity for mechanical quantities by overcoming the limitations of clamped resonators. However, the generally levitated silica nanoparticles (NPs) with low absorption and high transparence still face difficulties surviving in high vacuum with unclear reason. By monitoring the physicochemical properties like scattering, mass and density of amorphous silica NPs during pumping process. we propose that the loss of NPs may arises from the motional instability induced by laser heating lead releasing at low pressure. In this work, two types of NPs are heat treated from 100℃ to 1200℃ to release impurities before being loaded into an optical trap. The high vacuum levitation ratio for both NPs increase obviously after heat treatment. In particular, for NPs heated to 600℃, the ratio strikingly improves from ~35% to 100% and ~0 to 85% for two types of NPs. The loss mechanism is further confirmed by their relatively stable physicochemical parameters during pumping process. This work paves a way for wide application of levitated nano-resonators and indicates that levitated vacuum optomechanical systems could be a promising tool for dynamics and in-situ studying of small particles like aerosols and dusts.

**Introduction**

Nanomechanical resonators has significantly promoted the sensitivities for mechanical quantities such as force, mass, and charge, since the advent of atomic force microscopy [1-3]. Generally, their sensing abilities are ultimately limited by thermal noise and mechanical damping rates, hence clamped resonators usually work in cryogenic environment [4]. Levitated optomechanical system in vacuum that provide an excellent non-contact oscillator for precise measurements have attracted wide attention by solving limitations of clamped resonators[5-10].

Levitated optomechanical system in vacuum has high precision and advanced sensing performances for ultra-weak forces [11] [12], interactions [12-15], tiny masses [2, 16, 17], charges [18-20], polarizabilities [21], electric and magnetic fields [22, 23]. Zepto-Newton force sensitivity achieved with levitated NP in pressure of less than $1 \times 10^{-8}$ mbar is demonstrated [24], nano-$g$ acceleration sensitivity is achieved at $\sim 10^{-7}$ mbar [17, 25] and ultrasensitive torque sensitivity in the order of $10^{-27}$ Nm/$\sqrt{Hz}$ is achieved at $\sim 10^{-6}$ mbar, with sensitivity promotions of about three orders comparing with the most sensitive sensors then[13]. All the high sensitivities above rely on high vacuum environment, which poses challenges for realistic applications. Generally, NPs are loaded into optical traps using a nebulizer at ambient condition [26-28]. One big obstacle arises from the loss of NPs while pumping surrounding air to low pressure [29-31]. Although laser heating is ascribed for the NP loss, the complete loss mechanism is blurring and the solution for stable optical levitation is vacant. Precise in-situ physicochemical characterization of the trapped NPs may help lifting the veil.

In this work, the loss mechanism of NPs in optical trap is studied through its physicochemical alternation resulting from morphological changes during evacuation process. Further, thermal morphological tuning process is applied to NPs before loading them into traps to realize high ratio of stable levitation. At first, the loss pressure of NPs for both kinds of NPs are investigated. And the scattering light power and resonance frequency decrease of levitated NPs is studied along with the evacuation process, which reveal changes of NP's physicochemical properties. Then, the physicochemical properties like mass, size and density are carefully measured based on harmonic electric field driving method. Mass decrease of 26% for the NP without heat treatment reveals obvious material releasing. And the relatively small decrease of radius and increase of density indicate releasing from NP surface. To demonstrate, heat treatment is conducted on NPs before loading them into optical trap to get rid of redundant materials on the silica NPs. Herein, two types of amorphous silica NPs from two companies are used. For nearly all NPs, the high vacuum levitation ratio (HVLR) improves after heat treatment. The optimization of heat treatment especially works at 600℃, where the HVLR respectively improves from ~30% to 100% and 0 to 85% for two types of NPs. For NPs heated to 600℃, the changes in mass, density and radius are no more than 15%. The loss mechanism and thermal morphological tuning effects of NP is also confirmed by reduction of Si-OH groups and amplification of Si-O-Si featured by attenuated total reflectance Fourier transform infrared (ATR-FTIR) spectroscopy with increasing heat treatment temperature. The pre-heating treatment that effectively increase HVLR will enhance wide application of levitated nano-resonators and optomechanical systems could be a promising tool for researches on small particles like aerosols and dusts [32-35].

## Results

**NP lost during evacuation process at pressure of 10-0.1mbar.** Amorphous silica ($SiO_2$) NPs made by Stöber method from NanoCym company (sample A) and micro particles GmbH (sample B) are used in this work. Fig.1a schematically shows the trapping of NP in an optical trap. NPs are loaded into the trap at ambient pressure by nebulizing a solution of isopropyl alcohol (IPA) and monodispersed silica particles into the chamber. After a single NP is captured, the chamber pumping process starts to evacuate air molecules and solvents.

It is found that not all $SiO_2$ NPs fabricated by the Stöber method could withstand high vacuum pumping process. Statistical results on the HVLR and loss pressure of single-silica NPs from two companies are shown in Fig.1b. Single NP stable levitation tests are executed for 20 times for both kinds of NPs. $SiO_2$ NPs captured at ambient pressure tends to lose at pressure ranging from 10 to 0.1 mbar. Generally, NPs can afford high vacuum if they can surpass pumping to 0.1 mbar. For sample A, about 30% of the trapped NPs can survive high vacuum, while nearly no sample B could afford high vacuum.

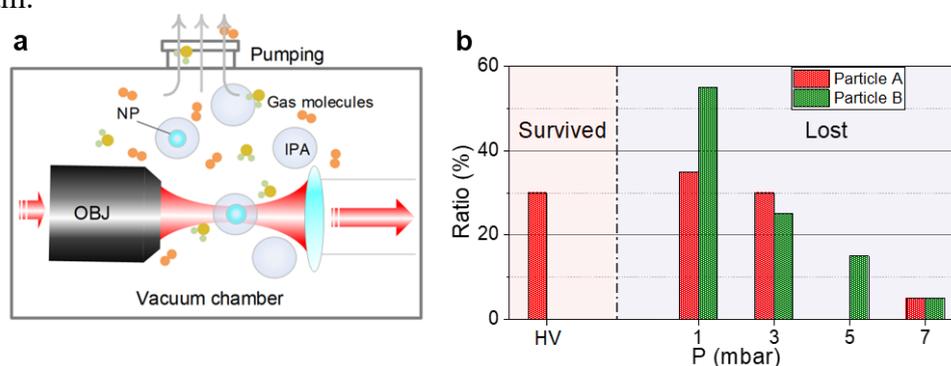

**Figure 1:** Capturing and loss of NPs from an optical trap. (**a**) Schematic of NP capturing setup. The chamber is pumped to low pressure after a single NP being captured. (**b**) The statistical loss pressure and high vacuum levitation ratio (HVLR) for NPs from two companies. Each kind of NP is tested for 20 times.

**Physicochemical characterization of NPs through evacuation process.** To explain the NP loss mechanism, the dynamic scattering light and resonant frequency for trapped NPs are monitored during the chamber evacuation process (Fig 2). Fig. 2a shows that the scattering light intensity and resonant frequency of sample A decrease from 1bar to 1mbar, then abruptly shrinking from 1mbar to 0.1 mbar, and then keeps constant at lower pressure.

To further understand, the dynamic changes of a NP are recorded by circulating the pressure from air condition to 50mbar, 20mbar, 10 mbar, 1mbar, and ~$10^{-2}$ mbar. Fig. 2b, c respectively shows the dynamic changes of a high vacuum non-sustainable and a high vacuum sustainable NP. The resonant frequency is only attainable at low damping condition, where the frequency decrease with decreasing pressure. And the scattering light intensity circulating back to approaching initial circulating pressure of more than 10 mbar. However, it can hardly recover after experiencing 1 mbar, and the scattering light keeps at low level after experiencing 0.01 mbar. This indicates that non-recoverable particle property change happened after circulating pressure of 10 mbar. The essential property changes of NP at around 1 mbar may arise from incremental thermal effects from light absorption at low air pressure, along which the particle motion drastically increase, resulting in the instability and loss of particles from the optical trap [36].

Fig. 2b shows the parameter changes of a NP lost at about 1.8 mbar during evacuation process. The loss pressure is located within the steeping change process. For a NP levitated with constant optical trap, both its scattering light and resonant frequency are determined by the polarizability $\alpha_p$ and mass m, corresponding to refractive index $n_r$, volume $V$ and density $\rho$ (see Supplementary Note 1). Thus, it is reasonable to correlate the loss of the NP to drastic change of the physicochemical properties of the NP.

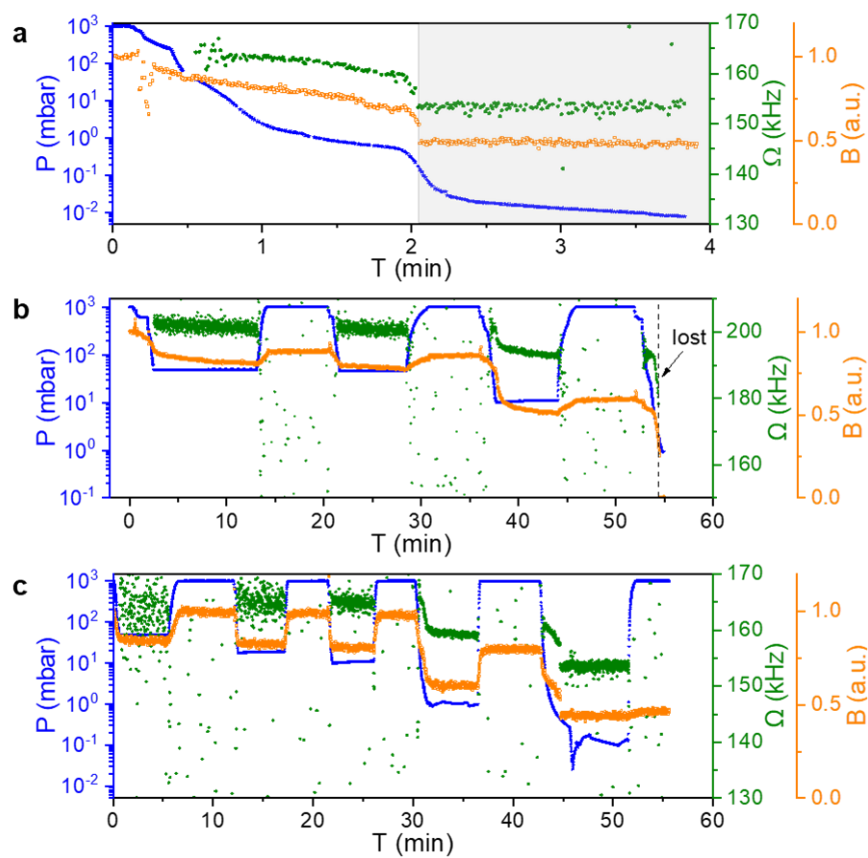

**Figure 2.** Scattering light and resonance frequency of trapped sample A measured through evacuation process. (**a**) Both scattering light power and resonance frequency slowly decrease with pumping process from atmosphere pressure to about 1mbar; then steeply decrease to about 0.1mbar; and finally keeps constant with pumping process. Scattering light power and resonance frequency of (**b**) a high vacuum non-sustainable and (**c**) a high vacuum sustainable NP measured with surrounding pressure cyclically changed. The resonance frequency cannot be extracted at atmosphere pressure due to overdamping. The scattering light power circulates back to high power before steep-decreasing process, while it barely changes after steep-decreasing. The NP that is not sustainable to high vacuum abruptly lost during steeping change process at 1.8 mbar.

**Precise mass, radius and density measurements of in-situ levitated NPs.** The vacuum pressure is level circulated between $P = 50$ mbar and decreasing to P=20, 10, 1, 0.1, and below 0.001mbar, as shown in Fig. 3a blue dashed line. The measurements are executed at each 50 mbar. Firstly, the mass of the particle P1 of sample A is detected through its respond to electric driving field[2]. Fig. 3b shows the power spectral density (PSD) of a thermally and harmonically driven resonator, where the NP's mass can be calculated by comparing the harmonically driven PSD to its thermal driven counterpart. Then, the radius $R$ of the trapped NP is calculated through its damping rates according to the kinetic theory of particles in gas [26,37]. Finally, the density of NP can be calculated by $\rho = m/V$ (see Supplementary Note 2).

The relative mass change of about 26% of NP P1 reveal non-negligible releasing of material of particle. The reduction of particle's radius and increase of density indicate material releasing from particle surface. And the shrink reduction from 10 mbar to 0.1mbar coincides with the frequently particle loss pressure. It is reported that the temperature of the trapped NP abruptly increase from room temperature to ~200 °C from10-0.1mbar[36]. For understanding, the internal temperature of a silica NP with diameter of 150 nm levitated in our system is calculated[38] (see Fig. S3). is calculated According to Zhuravlev model, which is widely used for surface analysis of amorphous silica, silica tends to go through dehydration and dihydroxylation process when temperature arise from room temperature to ~200 °C [39,40]. From the results above, it can be inferred that the chemical groups releasing during dehydration and dihydroxylation process result in force imbalance and lead to levitation instability.

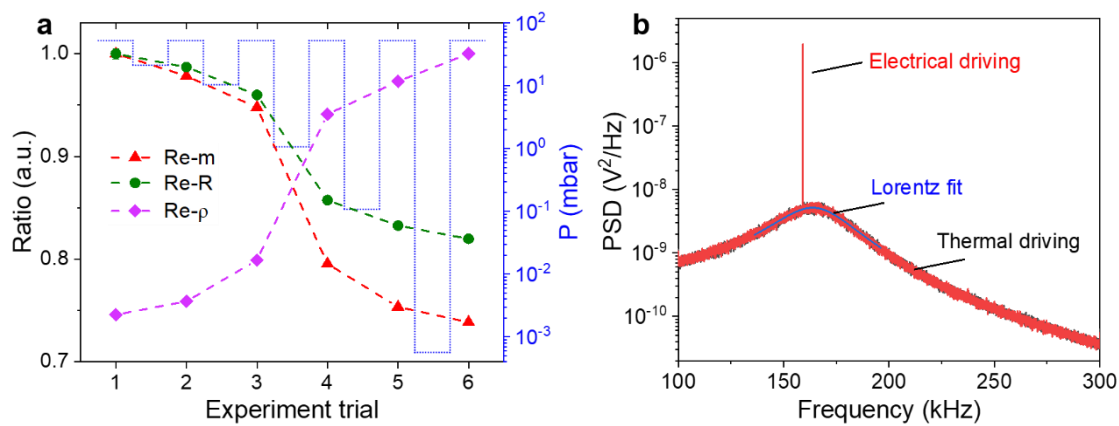

**Figure 3** Dynamic relative mass, radius and density measurements of a single NP. **(a)** The dynamic measurement is conducted at 50 mbar for each vacuum level circulations decreasing to P=20, 10, 1, 0.1, and below 0.001mbar. **(b)** The mass of a NP is calculated through its electric field driving part comparing with thermal actuation part.

**Enhancing HVLR through heating treatment.** To validate the above inference, heat treatment is conducted to change the properties of NPs before loading them into optical trap. To do this, the NPs are heated to high temperature including 100, 400, 600, 900 and 1200°C for comparison (see methods for detailed heating process).

Fig. 4a show the statistical results on the HVLR for sample A and B, in which half of sample A can surpass high vacuum evacuation and nearly none for sample B. It is obvious that the heat treatment to NPs is significantly beneficial to the levitation stability in high vacuum, especially for sample B. The HVLR increases with heating temperature ranging from 100-600°C, while decreasing when temperature is more than 600°C. The improvements especially work at heating temperature of 600°C, where all 20 samples of sample A and 17 to 20 samples of sample B surpassed high vacuum pumping process. Further, the dynamic physical parameters of a heated NP as an example are measured. Fig.4b show the mass, radius and density of particle P2 of sample A, which is pre-heated at 600°C. The mass, radius and density change of the pre-heated NP is remarkably smaller than that of untreated NPs.

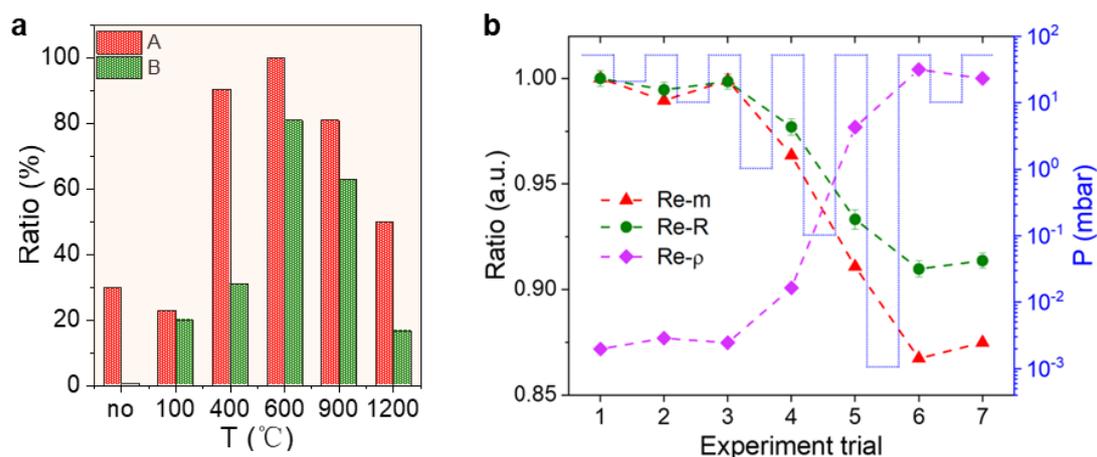

**Figure 4 Levitation performance of NP after heating treatment. (a)** Statistical results on HVLRs of two kind of NPs pre-heated at different temperatures. Each ratio is obtained from 20 independent tests. **(b)** Dynamic relative mass, radius, and density of a NP measured after heating treatment of 600°C.

**The physicochemical changes of NPs due to heat treatment.** To confirm the assumption of the dehydration and dihydroxylation process, attenuated total reflectance Fourier transform infrared (ATR-FTIR) spectra measurement of the heated and untreated NPs are executed. The morphological changes corresponding to temperatures in range of 100-1200 °C associated with the dehydration and dihydroxylation processes of silica NPs as indicated by the changes in Si-OH bond around 950 cm$^{-1}$, Si-O-Si bond around 795 cm$^{-1}$ and 1054 cm$^{-1}$ in the ATR-FTIR spectra in Fig. 5a. It is evident that the content of Si-OH bond reduces and Si-O-Si bond increase with raising heating temperature. This can explain the promoted HVLR of NPs with increased heating temperature.

To understand the decreasing HVLR at heating temperature of 900 °C and 1200 °C. The representative TEM images and morphological changes of sample A is measured as shown in Fig. 5b. Surface broken is found for NPs heated to 900 °C and 1200 °C. This may arise from strong stress induced by thermal expansion and contraction, which impairing the levitation stability of NPs. Fig. 5c ideally illustrates that the surface change of silica NPs with different heating temperature. As a result of heating to 600 °C, primary evaporation of IPA solvent and dehydroxylation of OH groups from silanol (Si-OH) bonds generate siloxane (Si-O-Si) bridges. Heat temperatures of 900°C and 1200 °C will partly result in surface broken.

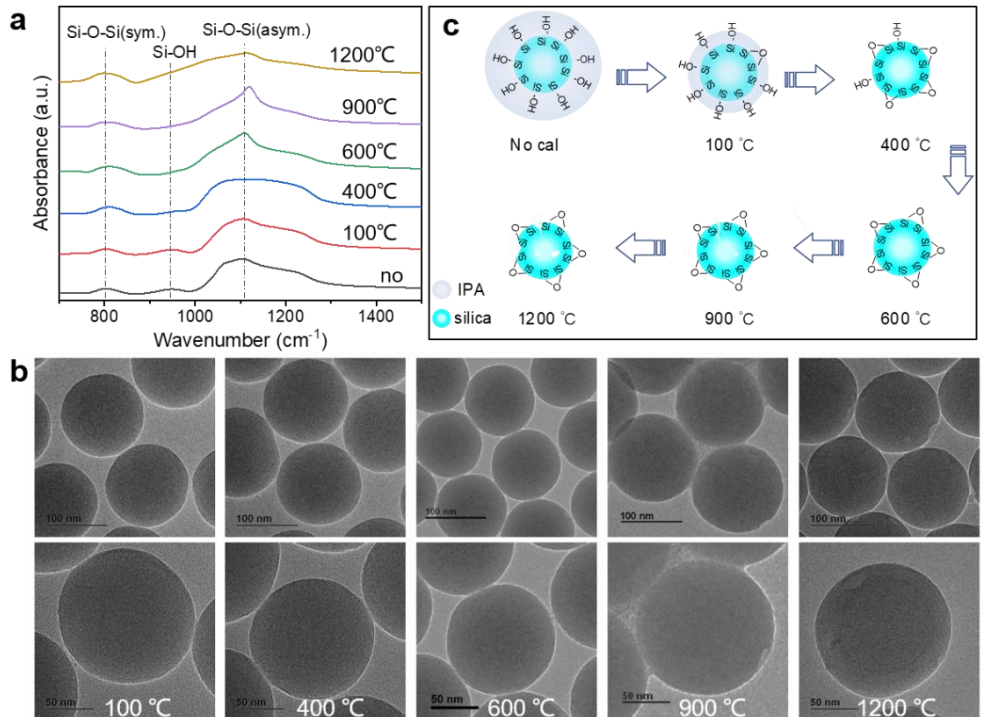

**Figure 5 Material changes of NP due to heating treatment. (a)** Changes in the Si-OH and Si-O-Si features for silica NPs pre-heated at different temperatures measured using attenuated total reflectance Fourier transform infrared (ATR-FTIR) spectroscopy. Bands around 795 cm$^{-1}$, 950 cm$^{-1}$, and 1054 cm$^{-1}$ correspond to Si-O-Si (sym), Si-OH, and Si-O-Si (asym) vibrations, respectively. With an increase in temperature, the Si-O-Si asymmetric band broadens while Si−OH vibration band ∼950 cm$^{-1}$ appears to be diminished in the heating-treated sample as a result of dehydroxylation of silica. **(b)** The morphological changes of sample A measured with transmission electron microscope (TEM) with two rows corresponding to two imaging scale. Surface broken is obvious for NPs heated to 900 °C and 1200 °C. **(c)** Schematic representation of chemo-morphological changes that occurring during sintering of silica NPs.

## Discussion

NP properties modifications are found by its decreasing scattering light intensity and resonance frequency. Material releasing from the surface of NPs in the process of evacuation is confirmed by the decreasing of mass, radius and density measured through its motion stimulated by alternating-electric fields. From the coincidental consistent NP loss and drastic modification pressure, an assumption that the air evacuation leads to the heat from laser absorption hard to diffusion and releasing of impurities from NP is proposed. The laser absorption aggravates the motion and flee of NPs from the optical trap.

To solve the problem, heat treatment is applied to get rid of impurities from amorphous silica NPs before loading them into optical traps. The levitation stabilities strikingly improve after heat treatment. The best HVLRs are obtained at heating temperature of 600°C, where the ratios respectively improved from ∼30% to 100% for sample A and∼0 to 85% for sample B. The impurity releasing is further confirmed by the relatively small mass, radius and density change of heated NP detected through evacuation process. The ATR-FTIR spectra of silica NPs indicate dehydration and dihydroxylation processes responsible for the heating releasing process.

**Conclusion**

In summary, we have proposed an explanation for NPs lost from optical trap and demonstrated a thermal tuning method to enhance the levitation stability of NPs in high vacuum. The high HVLR can be readily achieved by pre-heat treatment and boom the high precision sensing application of optical levitation systems in weak force, acceleration, electric field and novel physics. Further, more potential applications, such as the origin and evolution study on aerosol, moon and Martian dusts, can be realized through the precision dynamic characterization method on individual NPs.

**Methods:**

**1. Experimental setup**

The measurements were carried out using a home-built setup allowing silica nanospheres with radius of 50-150 nm to be optically trapped and monitored (see Supplementary Fig. S1. For NP trapping, a $\lambda$ = 1064 nm laser (ALS-IR-1064) was used with continuous-wave power before a high numerical aperture microscope objective (Nikon TU plan ELWD, 100×, NA = 0.8). Silica NPs from NanoCym company (SC21R-15RGX; diameter of 150 nm) and micro particles GmbH (SiO2-R-L3205-23; diameter of 142 ± 4nm) were used in this study. And they were high fold diluted into IPA before being loaded into the trap with a nebulizer (Omron, NE-C25S). The forward scattering beam from the trap was collected by an aspheric lens (Thorlabs AL1512-C, NA = 0.55), and then sent to a quad phase detection (QPD) (HAMAMATSU InGaAs PIN photodiode G6849) for the detection of the particle's center-of-mass motion along the three axes. A 3kV high voltage signal is applied to manipulate the charge of NPs through an electrode by ionization.

**2. NP heat treatment**

Firstly, NP suspensions are centrifuged and freeze-dried to extract NPs from solvents. Then, the NP heating process is performed in a KSL-1100X muffle furnace (HF-Kejing company). The samples are gradually heated to dedicated temperature with heating rate of 5℃/min, and then sintered for three hours. For heat temperature of less than 300℃, the muffle furnace is naturally cooled to room temperature; For heating temperature of more than 300℃, the muffle furnace is firstly cooled to 300℃ of 5℃/min, and then naturally cooled to room temperature. After heat treatment, the samples are diluted to IPA and dispersed through ultrasonic oscillation. For samples hard to disperse, Blender or Grinder is applied.

**3. Characterization of silica NPs through TEM and XRD**

To evaluate the changes in the chemical bonding during thermal treatment processes, infrared (IR) spectra were acquired in an Attenuated Total Reflection (ATR) mode using an attenuated total reflectance Fourier transform infrared spectrometer (ATR-FTIR, Nicolet iS 20, Thermo Fisher Scientific) for heated and untreated silica NPs (100, 400, 600, 900 and 1200 ℃). The spectra were collected in the range of 4000-650 cm$^{-1}$ with total number of 32 scans for each run.

**Acknowledgements**

This work was supported by the National Natural Science Foundation of China (42004154), Major Scientific Project of Zhejiang Laboratory (2019MB0AD01, 2020MB0AL03).

**Author contributions**
C.L. conceived the idea and wrote the paper. Y.M., J.W. prepared the setup and performed the experiments; S.Z. built the control and data acquisition systems; X.J. and H.Y. conducted the sample heat treatment; X.G. designed the data process method; Z.F. designed the electric control system; H.H. supported and supervised the joint efforts. All authors discussed the results and participated in writing the manuscript.

# Supplementary Information

This file includes:
Supplementary Note 1, Supplementary Note 2
Figure S1- Figure S4

**Supplementary Note 1** Characterization NPs through dynamic scattering light and resonance frequency

## 1. Particle information from scattering light power.

The polarization density for a small particle in a uniform electric field space with strength of $E_0$ is

$$P_d = \chi_e \varepsilon_0 E = \eta \chi_e \varepsilon_0 E_0, \qquad (1)$$

where $\chi_e$ is electric susceptibility of particle, $\varepsilon_0$ is dielectric constant of air and $E$ is electric field strength in the particle which is dependent on particle shape with $E = \eta E_0$. Suppose the volume of the NP is $V$, its electric dipole moment is

$$p = P_d V = \eta \chi_e \varepsilon_0 V E_0 = \alpha_p V E_0. \qquad (2)$$

Based on Rayleigh approximation for light scattering, a NP irradiated by a laser beam can be treated as an electric polarizability, whose electric dipole moment oscillates at the same frequency as the electric field component of the optical field:

$$p(t) = \alpha_p V E_0 \cos(\omega t). \qquad (3)$$

According to classical electrodynamics, dipole oscillations radiate electromagnetic waves of the same frequency. This radiation field can be considered as the scattered light field of the particle to the incident light. At the far field, the average energy flow density of the oscillating dipole far field radiation is

$$\bar{S} = \frac{\omega^4 p^2}{32\pi^2 \varepsilon_0^2 c^3 R^2} \cos^2\theta, \qquad (4)$$

where $\theta$ is light scattering angle. Thus, the Rayleigh scattering light intensity distribution of a NP irradiated by a linearly polarized light can be expressed as

$$I(V, \theta, L) = I_0 \frac{\alpha_p^2 V^2}{16\pi^2 \varepsilon_0^2 L^2} \left(\frac{2\pi}{\lambda}\right)^2 \cos^2\theta. \qquad (5)$$

Here, $L$ denotes the distance to the particle. Obviously, the scattering light intensity is linearly related to the square of particle volume. Particle size only exerts relatively small influence on scattering light. For simplicity, the volume of a NP can be derived by assuming a homogeneous sphere. The Rayleigh scattering light intensity distribution can be expressed as

$$I(V, \theta, L) = I_0 \frac{9\pi^2 V^2}{\lambda^4 L^2} \left(\frac{n^2-1}{n^2+1}\right)^2 \cos^2\theta. \qquad (6)$$

It's obvious that the scattering light power of a NP relies on its volume and refractive index.

## 2. Particle information from resonant frequency.

For a NP trapped in an optical trap, the stiffness of the resonator along three axes are respectively:

$$k_{x,y} = \frac{4\alpha' P_0}{\pi \varepsilon_0 c \omega_{0x} \omega_{0y}} \frac{1}{\omega_{0x,y}^2}, \qquad (7)$$

$$k_z = \frac{16\alpha' P_0}{\pi \varepsilon_0 c \omega_{0x} \omega_{0y}} \frac{1}{\omega_{0z}^2}, \qquad (8)$$

where $P_0$ is the laser power, $\omega_{0x/y/z}$ denotes beam radius along $x/y/z$ axis, $\varepsilon_0$ is the permittivity of air. The polarizability $\alpha' = 4\pi\varepsilon_s r^3 \frac{\varepsilon_p - \varepsilon_s}{\varepsilon_p + 2\varepsilon_s}$, where $\varepsilon_s$ and $\varepsilon_p$ are permittivity of surrounding environment and trapped particle. The resonant frequency for the motion along three axis are $\Omega_{x,y,z}=\sqrt{\frac{k_{x,y,z}}{m}}$, with $m=\frac{4}{3}\pi r^3 \rho$. Thus, it can be deduced that the resonant frequency changes with particle density $\rho$ and permittivity $\varepsilon_p$.

**Supplementary Note 2** The mass, density and radius measurement of NPs.

The motion equation of a particle with mass $m$ and the number of charge $n_q$ captured in an optical trap can be described by a thermally and harmonically driven damped resonator[2]:

$$\ddot{x}+\Gamma\dot{x}+\omega_0^2 x = \frac{F_{\text{th}}(t)}{m} + \frac{n_q q_e E_0 \cos(\omega_{\text{dr}} t)}{m}. \tag{9}$$

Here, $\Gamma$ is the damping rate, with $F_{\text{th}}(t)$ being random collisions with residual air molecules in the chamber. $\omega_0$ is the mechanical eigenfrequency of the oscillator. $\omega_{\text{dr}}$ is the frequency of the electric field driving response signal. $q_e$ is the elementary charge. $E_0$ is the electric field strength. The overall power spectral density(PSD) of a thermally driven mechanical resonator subject to external harmonic force reads:

$$S_x(\omega)=S_{\text{th}}(\omega)+S_{\text{dr}}(\omega) = \frac{2k_B T \Gamma}{m[(\omega_0^2-\omega^2)+(\Gamma\omega)^2]} + \frac{n_q^2 q_e^2 E_0^2}{2m^2}\frac{\tau \text{sinc}[(\omega-\omega_{\text{dr}})\tau]}{(\omega_0^2-\omega^2)+(\Gamma\omega)^2}, \tag{10}$$

where $k_B$ is the Boltzmann constant and $T$ is room temperature. $S_{\text{th}}(\omega)$ is the thermally driven (PSD), and $S_{\text{dr}}(\omega)$ is electrically driven power spectral density. $\tau$ is measurement time.

According to the relationship between the mass of particle, the electric force and PSD signal in the theoretical model of electrical drive of the particle levitated within optical trap, the relationship between mass of sample And net charge on a particle is obtained:

$$m = \frac{n_q^2 q_e^2 E_0^2 \tau}{8 k_B T \Gamma R_s}, \tag{11}$$

Where $R_S$ is the ratio of power spectral density of electric field driving and thermal driving parts, which can be expressed as $R_s = S_{\text{dr}}(\omega)/S_{\text{th}}(\omega)$. According to the theory of gas dynamics, in the thermal equilibrium environment, the damping rate of a NP with radius of $R$ and mass of $m$ can be expressed as:

$$\Gamma_0 = \frac{6\pi \eta_{air} R}{m}\frac{0.619}{0.619+k_n}(1+c_k). \tag{12}$$

Here, $c_k=0.31k_n/(0.785 + 1.152 k_n + k_n^2)$, $\eta_{air}$=1.82×10⁻⁵Pa·s is the air viscosity and $k_n = \frac{\bar{l}}{R}$ is the Knudsen number. The term $\bar{l} = k_B T/\sqrt{2}\pi d^2 p_{gas}$ is the mean free path of air molecules, $d = 0.353\times10^{-9}$ m is the diameter of air molecular. $p_{gas}$ is the ambient pressure. $\Gamma_0$ can be obtained by fitting $S(\omega)$ with the least square method. From the radius of the particle $R$, we can obtain the particle density $\rho = 3m/4\pi R^3$.

**Error analysis**

To determine the error in the calculation, a careful study of all the sources of errors has to be carried out based on the existing parameter conditions. For several variables and constants, we can neglect the corresponding uncertainty. $q_e$, $k_B$ and $\tau$ are all constants, and $R_s$ is calculated from the measured $S_{dr}(\omega)$ and $S_{th}(\omega)$, which are only affected by statistical errors. The error(thermal equilibrium) of mass of is given by:

$$\frac{\Sigma_m}{m} = \sqrt{(2\frac{\sigma_{n_q}}{n_q})^2 + (2\frac{\sigma_{E_0}}{E_0})^2 + (2\frac{\sigma_\Gamma}{\Gamma})^2 + (\frac{\sigma_T}{T})^2 + (\frac{\sigma_{R_s}}{R_s})^2 + (\frac{\sigma_\tau}{\tau})^2 + (\frac{\sigma_{k_B}}{k_B})^2}. \quad (12)$$

The uncertainty of various parameters that cause errors in thermal equilibrium in the experiment is shown in the following table for a NP repetitively measured for 19 times as shown in Fig. S3:

**Table1** Uncertainties table. Black font indicates errors that cannot be ignored. Gray font indicates that $\sigma_{z_i} \sim 0$.

| Quantity | Value $z_i$ | Error $\sigma_{z_i}/z_i$ |
|---|---|---|
| $T$ | 295.38 K | 2.4‰[a] |
| $\Gamma$ | 4.47×10³ Hz | 3‰[b] |
| $S_{th}(\omega)$ | 2.8171×10⁻⁹ V²/Hz | 2.4 ‰[c] |
| $S_{dr}(\omega)$ | 3.8716×10⁻⁷ V²/Hz | 2.6 ‰[c] |
| $E_0$ | 334 V/m | 1.1%[3] |
| $k_B$ | 1.380×10⁻²³ J/K⁻¹ | 5.72×10⁻⁷[5] |
| $q_e$ | 1.602×10⁻¹⁹ C | 6.1×10⁻⁹[5] |
| $\tau$ | 0.015 s | 1 ppm[d] |
| $n_q$ | 2 | 0 |
| $m$ | 1.964 fg | 2.32% |

[a] This error comes from temperature measurement error.
[b] This error comes from the error brought by the fitting process.
[c] This error comes from the measurement error caused by data fluctuation during data acquisition.
[d] Nominal value from the datasheet of lock-in amplifier (Zurich Instruments MFLI).

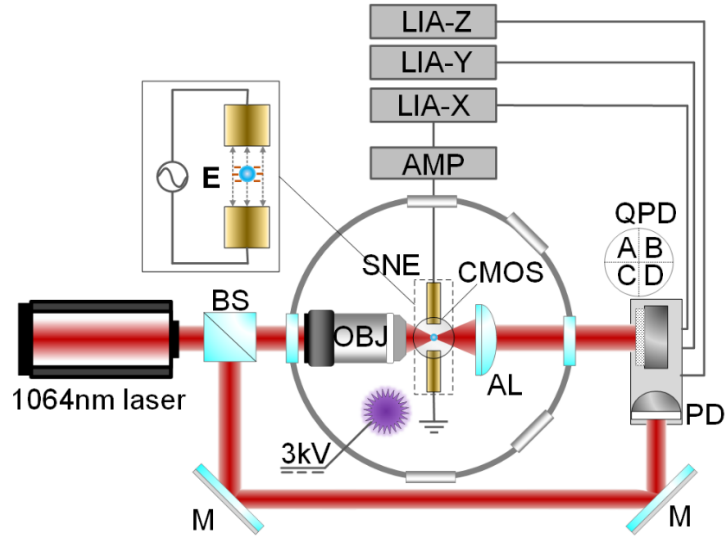

**Figure S1:** Experimental set up. The experimental setup used in the present study is upgraded from that used in Ref. [24]. The optical trap was created using a 1064 nm laser (ALS-IR-1064) beam and a high numerical aperture microscope objective (OBJ; Nikon TU plan ELWD, 100, NA=0.8). To detect the scattering light power of each captured particle, a CMOS beam profiler from DataRay was mounted vertically over the vacuum chamber. A self-developed photodetector that consists of a quadrant

photodetector (QPD; Hamamatsu G6849) and a normal photodetector (PD Hamamatsu G8370-82) is mounted for the detection of the particle's center-of-mass (COM) motion along the three axes. Three Lock in amplifiers (ZI-MFLI) are applied for signal acquisition. The forward scattering beam from the trap was collected by an aspheric lens (AL; Thorlabs AL1512-C, $f$ =12 mm, NA=0.55) and send to QPD, and a reference laser beam send to PD. A pair of electrodes composed of two horizontal steel (40CrMoV5) needles (SNE) that are 1 mm in diameter and placed 2.52 mm apart are mounted to produce electric field to control the motion of the trapped NP along control X axis. The electric driving signal from LIA-X is amplified by an AMP (Aigtek ATA-2031) and loaded onto the electrodes as a reference signal and the motion signals of X axes with the same frequency is extract out. Similar to Ref. [21], the FDTD numerically simulated value of electric intensity is employed. The purple glow on the side of the chamber is emitted by a wire connected to a high voltage (HV) DC source and is used to control the net charge of the particle. BS: beam splitter; M: mirror.

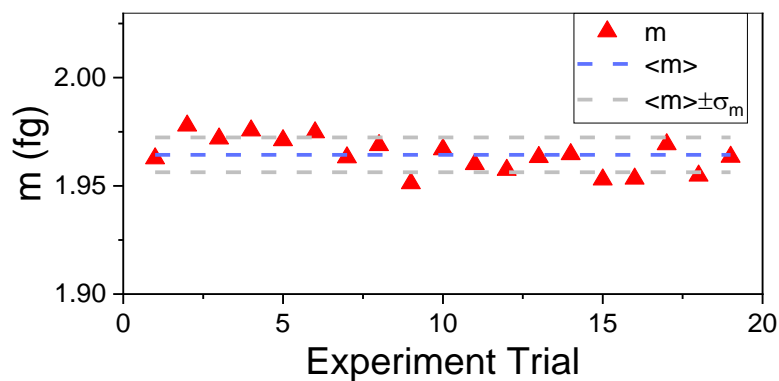

Figure S2: The mass measurement consistency of the harmonic driven method. The mass is measured at for 19 times for a NP of sample A. The statistical error is maintained below 1%.

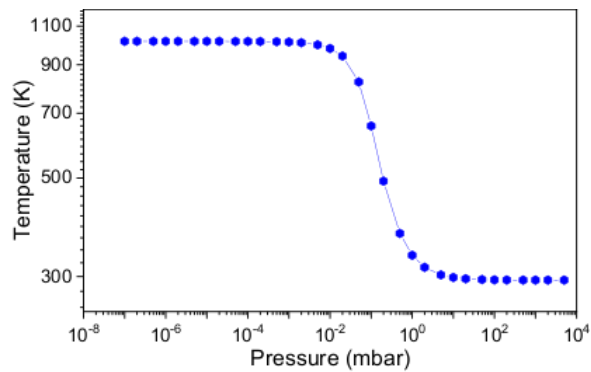

Figure S3: The calculated internal temperature of the trapped silica NP with diameter of 150 nm changes with pressure ranging from 104 mbar to 10-7 mbar. Laser absorption, thermal conduction and black body radiation are considered (Ref. [38]).

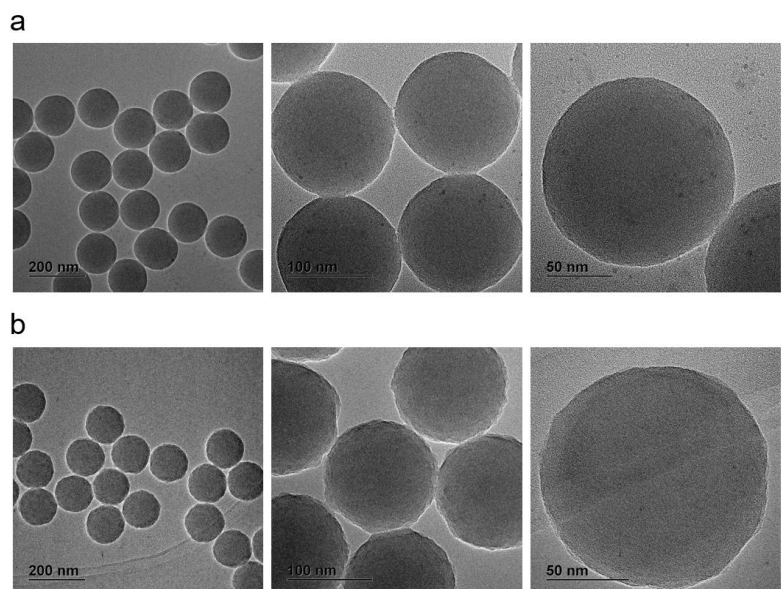

Figure S4: TEM images of two types of silica NPs from **(a)** NanoCym company (Sample A) and **(b)** micro particles GmbH (Sample B). The radius for two kinds of NPs are $71.4 \pm 3.6$ nm and $68.5 \pm 1.1$ nm.

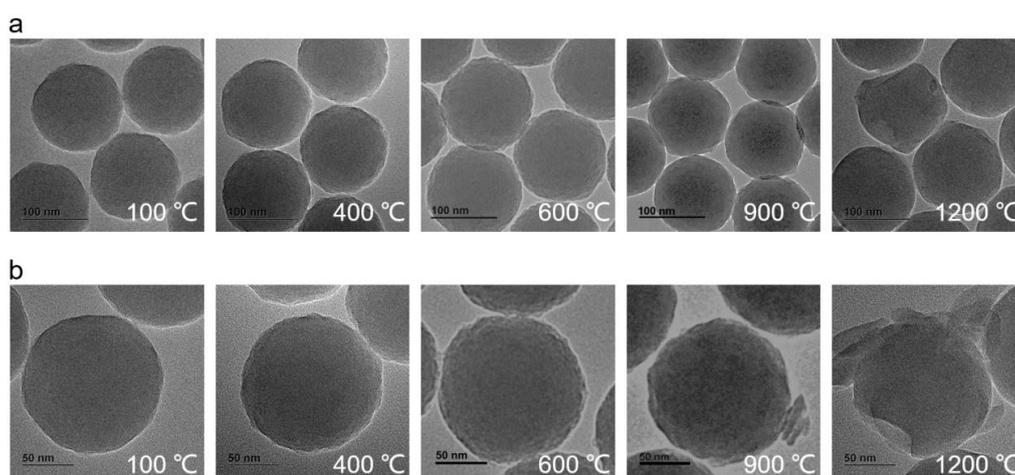

Figure S5: TEM images of silica NPs of sample B heat-treated at different temperature. Surface broken and debris are obvious after heat treating temperature of 900°C and 1200°C.

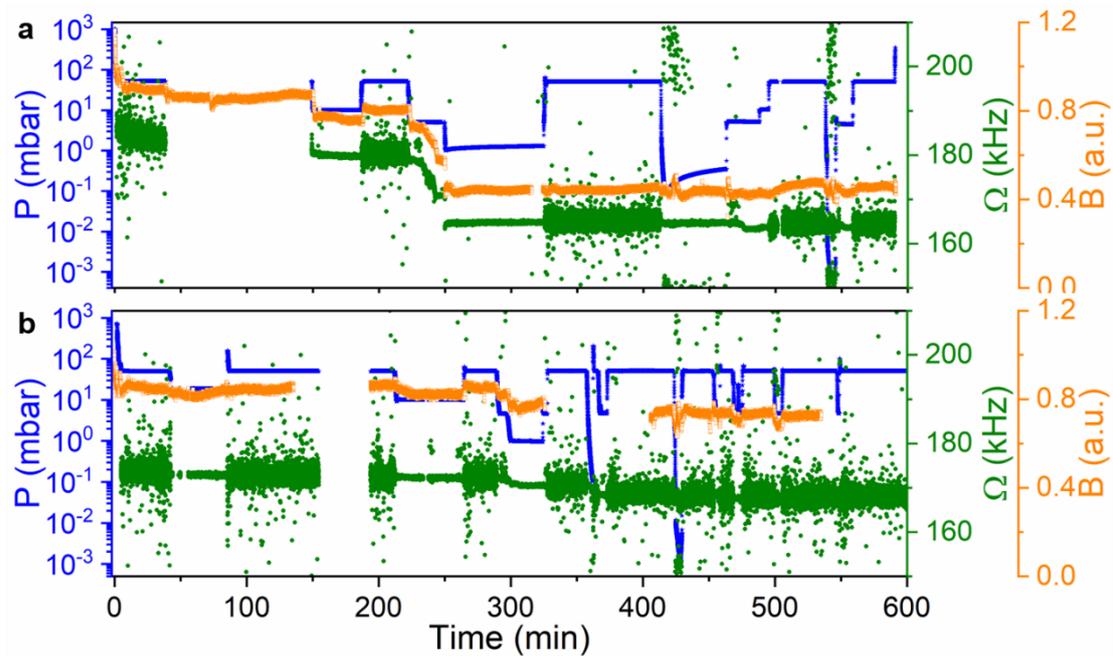

Figure S6: Scattering light power and resonant frequency monitored through pressure manipulation process for (a)P1(unheat-treated) and (b) P2 (600°C heat-treated).